\documentclass[aps,nofootinbib,showkeys,showpacs,preprint]{revtex4}
\usepackage{graphicx}
\usepackage{amsfonts}
\usepackage{subfigure}

\begin{document}

%\preprint{hep-th/yymmnnn}
\vspace*{3cm}
\title{Note on deconfinement temperature with chemical potential from AdS/CFT}

\author{Wen-Yu Wen} \email{steve.wen@gmail.com}
\affiliation{Department of Physics and Center for Theoretical
Sciences, National Taiwan University, Taipei 106, Taiwan}

%\date{\today}

\begin{abstract}
In this note we study the first-order Hawking-Page phase
transition of R-charged black hole in {\it truncated} Anti-de
Sitter space of various dimensions.  This corresponds to
confinement/deconfinement phase transition in the dual gauge field
theory with fixed chemical potential.  We demonstrate in general
this critical temperature decreases with increasing charge density
but with decreasing IR cutoff.
\end{abstract}



\maketitle

\section{Introduction}

Anti-de Sitter space-Conformal Field Theory (AdS/CFT)
correspondence has been widely studied since the work by
\cite{Maldacena:1997re,Gubser:1998bc,Witten:1998qj}. In
particular, its application to relate the thermodynamics of ${\cal
N}=4$ super Yang-Mills (sYM) theory in four dimensions to the
thermodynamics of Schwarzschild black holes in five-dimensional
Anti-de Sitter space\cite{Witten:1998zw}.  In this description,
confinement/deconfinement phase transition of gauge theory on a
sphere has its holographically gravitational description as the
Hawking-Page phase transition\cite{Hawking:1982dh}.  Later,
authors in
\cite{Klebanov:2000hb,Maldacena:2000yy,Polchinski:2000uf} realized
confinement by capping off the geometry (i.e. the Calabi-Yau cone)
smoothly at the infrared tip.  Though it has been known since then
that Hawking-Page transition happens at zero temperature for gauge
theory on a flat space, the papers
\cite{Boschi-Filho:2006pe,Herzog:2006ra,Ballon Bayona:2007vp}
showed that a finite Hawking-Page transition temperature can still
occur once an IR cutoff is introduced in AdS Poincar\'{e} patch.
This transition can be understood as first-order
confinement/deconfinement phase transition in the holographic
gauge field theory in the Minkovski space. Transition with matter
was recently studied in \cite{Kim:2007em}.  Transition up to
${\cal O}(\alpha'^3)$ correction was studied in \cite{Cai:2007bq}.
Phase transition temperature was obtained in relation to chemical
potentials for $N=4$ sYM on $S^3$ via direct evaluation of
partition function\cite{Harmark}.

In this note, we would like to extend this study to a more general
background.  To be specific, we will target at gauge theories with
fixed chemical potential in spacetime dimension $3,4,5$ and $6$,
via the study of a special kind of R-charged black hole in AdS
space of one dimension higher according to the AdS/CFT
correspondence\cite{Cvetic:1999xp}.

We will also introduce an IR cutoff essential to realize the
confinement/deconfinement phase transition and mimic a QCD-like
model. There are hard wall and soft wall models proposed in the
literature\cite{Polchinski:2001tt,Erlich:2005qh,Andreev:2006eh}
and our result shows that hard wall model is enough to catch new
features contributed from charge density and IR cut-off.

We begin in section II with brief review on the thermodynamics of
these R-charged black holes in general.  In section III we
evaluate the Gibbs Euclidean action and obtain a implicit formula
for the Hawking-Page transition temperature in various dimensions.
In section IV we restrict to the case of single charge for
simplicity and study the transition temperature in each specific
dimension. We find that in most cases the Hawking-Page transition
temperature decreases with increasing charge density, but with
decreasing IR cut-off.  We conclude with some comments in section
V.

\section{Thermodynamics of general R-charged black holes}

Although the supergravity theories vary in different spacetime
dimensions, here we consider a general bosonic action for gravity
coupled to a set of scalars and vector fields given in the form,
\begin{eqnarray}\label{action_reg}
I=&&-\frac{1}{16\pi G_d}\int_{\cal
M}{d^dx}\sqrt{-g}[R-\frac{1}{2}{\cal
G}_{ij}(\phi)\partial_\mu\phi^i\partial_\nu\phi^j-\frac{1}{4}G_{IJ}F_{\mu\nu}{}^IF^{\mu\nu}{}^J-V(\phi)]\nonumber\\
&&+\frac{1}{8\pi G_d}\int_{\partial {\cal
M}}d^{d-1}\sqrt{-h}\Theta.
\end{eqnarray}
This action is composed of a term integrated over $d$-dimensional
spacetime $\cal M$ and the other over a $(d-1)$-dimensional
boundary $\partial {\cal M}$. The latter is the Gibbon-Hawking
surface term given in terms of trace of the extrinsic curvature
$\Theta_{\mu\nu}$, defined as
\begin{equation}
\Theta_{\mu\nu}=-\nabla_{(\mu}n_{\nu)},
\end{equation}
where $n^\mu$ is the outward-going normal on $\partial {\cal M}$
and $h_{\mu\nu}$ is the induced metric. Having in mind that $\cal
M$ is chosen to be asymptotic $AdS_d$ space and $\partial {\cal
M}$ is its boundary with topology $R_t\times S^{d-2}$, we are
interested in spherically symmetric black holes carrying electric
charge(s), i.e.
\begin{eqnarray}
&&ds^2=-e^{2(d-3)B(r)}f(r)dt^2+e^{2B(r)}(\frac{dr^2}{f(r)}+r^2d\Omega_{d-2}^2),\nonumber\\
&&\phi^i=\phi^i(r),\qquad A_t{}^I=A_t{}^I(r).
\end{eqnarray}
For a black hole solution, we request finite $B(r)$ and vanishing
$f(r)$ at the event horizon $r=r_+$, which is the largest real
root satisfying equation $f(r)=0$. The asymptotic geometry is
obtained by sending $r\to \infty$ so that
\begin{equation}
\lim_{r\to\infty}B(r)= 0, \qquad \lim_{r\to\infty}f(r)= g^2r^2,
\end{equation}
where $g$ is the inverse of AdS radius.  To our interests, a class
of stationary R-charged black hole shares a common
description\cite{Cvetic:1999xp}, i.e.
\begin{eqnarray}
&&e^{2(d-2)B(r)}\equiv {\cal H}(r)=\prod_{I=1}^n{(1+\frac{q_I}{r^{d-3}})},\nonumber\\
&&f(r)=1-\frac{\mu}{r^{d-3}}+g^2r^2{\cal H}(r),
\end{eqnarray}
where $n$ is the maximum number of $U(1)$ which can be embedded
inside the R-symmetry group for each specific dimension.  This
black hole can be viewed as Kaluza-Klein reduction of a rotating
black brane on the sphere, after taking certain
limit\cite{Cvetic:1999xp,Cvetic:1999ne}.  The angular momentum in
the full ten or eleven dimensions plays the role of R-charge, i.e.
$U(1)$ charge to the gauge field $A^I$ in the $d$-dimensional
effective theory.

Now we will follow \cite{Batrachenko:2004fd} for the discussion of
its thermodynamics, but also see
\cite{Chamblin:1999tk,Cvetic:1999ne}.  The appropriate ensemble
applied to that with a nontrivial chemical potential is the
grand-canonical ensemble, and the on-shell action
(\ref{action_reg}) relates to the Gibbs free energy by $I=\beta
\Omega$, where
\begin{equation}\label{gibbs}
\Omega = E-TS-\Phi^IQ_I.
\end{equation}
This relation needs some explanation.  $\beta$ is the periodicity
of Euclidean time and inverse of Hawking temperature in thermal
equilibrium, i.e.
\begin{equation}\label{temp}
\beta^{-1}=T=\frac{1}{4\pi}e^{-(d-2)B(r)}f'(r)\vert_{r_+}.
\end{equation}
The entropy can be obtained via the Bekenstein-Hawking relation,
\begin{equation}
S=\frac{\omega_{d-2}}{8\pi G_d}(2\pi
e^{(d-2)B}r^{d-2})\vert_{r_+},
\end{equation}
where $\omega_{d-2}$ is the volume of $S^{d-2}$.  $E$ is the black
hole mass , here given by
\begin{equation}\label{mass}
E=<T_{tt}>=\frac{2}{\sqrt{-h}}\frac{\delta \Omega}{\delta
h^{tt}}=\frac{\omega_{d-2}}{8\pi
G_d}\sqrt{-h}(-\Theta^t{}_t+\Theta).
\end{equation}
At last, the chemical potential $\Phi^I$ and the normalized charge
$Q_I$ are given by
\begin{eqnarray}
&&\Phi^I=A_t{}^I(\infty)-A_t{}^I(r_+),\nonumber\\
&&Q_I=\frac{\beta\omega_{d-2}}{16\pi G_d}q_I.
\end{eqnarray}
The only subtlety is that the action (\ref{action_reg}) and mass
(\ref{mass}) are still divergent though it has been regularized
via the Gibbon-Hawking term.  The paper \cite{Batrachenko:2004fd}
demonstrates that a holographic renormalization scheme naturally
applies to render finite quantities but also points out that
(\ref{gibbs}) is still valid regardless of renormalization.  In
this note we only concern the difference between free energy of
black hole and that of thermal AdS, and expect those counterterms
cancel each other. Therefore, the regularized action is enough to
our purpose\footnote{However, for the case of multiple charges or
higher dimensions, we may need a counterterm at IR cut-off thanks
to the truncated AdS space.  We will deal with this subtlety later
we face it.}.

\section{Hawking-Page analysis and deconfinement phase transition}
In this section, we are going to determine the phase transition
temperature via the Hawking-Page analysis adopted in
\cite{Herzog:2006ra}.  To proceed, we impose an infrared cut-off
at $r=r_c$ as in the hard wall model proposed in
\cite{Erlich:2005qh} as well as an ultraviolet cut-off at $r=r_0$.
The IR cut-off is essential for confinement.  The UV cut-off is
only for calculation convenience and will be sent to infinity at
the end.  We consider the following three on-shell actions:

\begin{equation}
I_1=\frac{\beta\omega_{d-2}}{8\pi
G_d}[(3-d)r_0^{d-3}f(r_0)-\frac{1}{2}r_0^{d-2}f'(r_0)-r_0^{d-3}+r_+{}^{d-3}],
\end{equation}
for $AdS$ black hole where $r_+>r_c$.
\begin{equation}
I_2=\frac{\beta\omega_{d-2}}{8\pi
G_d}[(3-d)r_0^{d-3}f(r_0)-\frac{1}{2}r_0^{d-2}f'(r_0)-r_0^{d-3}-r_c^{d-2}f(r_c)B'(r_c)-r_c^{d-3}(f(r_c)-1)],
\end{equation}
for $AdS$ black hole where $r_+<r_c$.  At last for thermal AdS
with non-vanishing gauge field, we have
\begin{equation}
I_3=\frac{\beta'\omega_{d-2}}{8\pi
G_d}[(3-d)r_0^{d-3}f_0(r_0)-\frac{1}{2}r_0^{d-2}f'_0(r_0)-r_0^{d-3}-r_c^{d-3}(f_0(r_c)-1)],
\end{equation}
where
\begin{equation}
f_0(r)\equiv\lim_{\mu\to0}f(r)=1+g^2r^2{\cal H}(r).
\end{equation}
We request asymptotically two geometries have the same periodicity
along Euclidean time , say
$\sqrt{f_0(r_0)}\beta'=\sqrt{f(r_0)}\beta$. Then the difference of
Gibbs energy, after the UV cut-off removed, becomes
\begin{equation}\label{V13}
\Delta V_{13}\equiv
-\lim_{r_0\to\infty}\beta^{-1}(I_1-I_3)=\frac{\omega_2}{8\pi
G_d}[\frac{\mu}{2}-r_+^{d-3}-g^2r_c^{d-1}{\cal H}(r_c)].
\end{equation}
The indefinite sign of $\Delta V_{13}$ indicates a first-order
Hawking-Page transition at $\Delta V_{13}=0$.  It is energetically
preferred for thermal AdS with non-vanishing gauge field at low
temperature, but R-charged black hole at hige temperature. When
the black hole is unaccessible in the truncated AdS, i.e.
$\infty>r>r_c>r_+$, the difference becomes
\begin{equation}
\Delta V_{23}\equiv
-\lim_{r_0\to\infty}\beta^{-1}(I_2-I_3)=\frac{\omega_2}{8\pi
G_d}[\frac{\mu}{2}+r_c^{d-2}f(r_c)B'(r_c)]
\end{equation}
The positivity of $\Delta V_{23}$ indicates that thermal AdS with
non-vanishing gauge field is always energetically preferred
whenever black hole is unaccessible in the truncated AdS space.

\section{Deconfinement phase transition in various dimensions}
Now we are ready to investigate the confinement/deconfinement
phase transition in various dimensions via the above Hawking-Page
analysis on R-charged black hole in the AdS space.  For
simplicity, we will restrict our discussion to the case of single
R-charge, say $q_1=q$ and $q_J=0$ for $J\neq 1$. Equation
(\ref{V13}) states that the Hawking-Page transition happens at
\begin{equation}
\frac{\mu}{2}-r_+^{d-3}-g^2r_c^{d-1}{\cal H}(r_c)=0,
\end{equation}
and one can in principle inverse equation (\ref{temp}) to relate
$r_+$ to the deconfinement transition temperature $T_{c}$, though
the resulting polynomial equation may not be solved analytically.

\subsection{$4d$ black hole/thermal $3d$ super Yang-Mills}
\begin{figure}
\begin{flushleft}
     \subfigure[Hawking-Page analysis of AdS$_4$ R-charged black hole. The origin represents zero
charge density without IR cut-off where $T_c=\frac{g}{\pi}$.]{
          \label{fig1}
          \includegraphics[width=.45\textwidth]{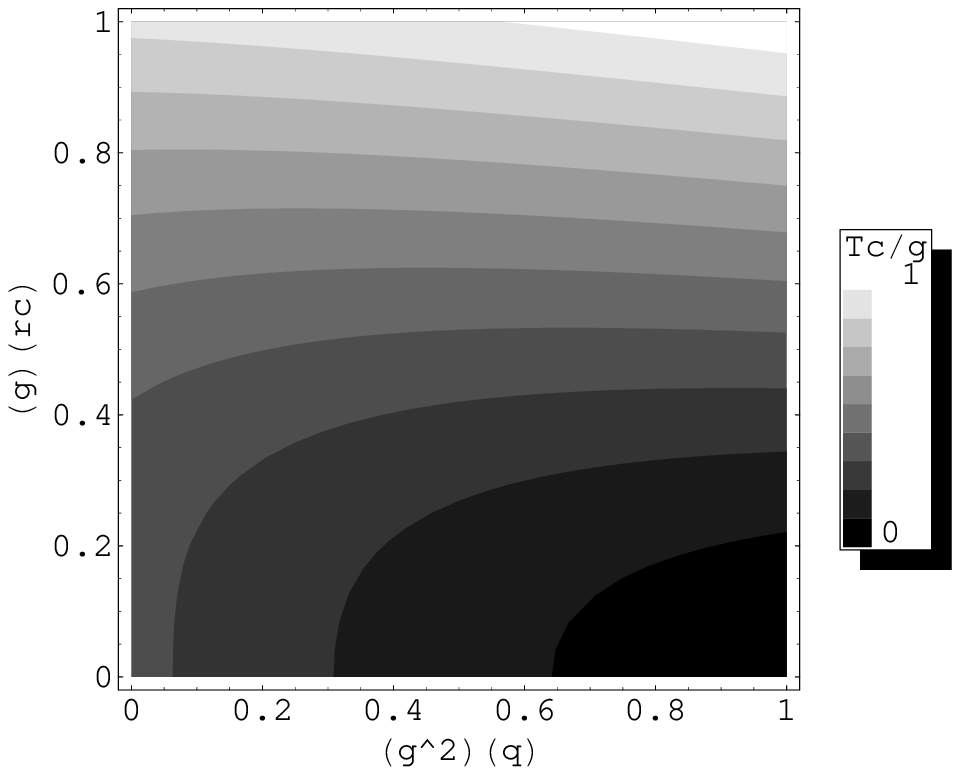}}
     \hspace{.3in}
     \subfigure[Hawking-Page analysis of AdS$_5$ R-charged black hole. The origin represents zero
charge density and no IR cut-off where $T_c=\frac{3g}{2\pi}$.]{
          \label{fig2}
          \includegraphics[width=.45\textwidth]{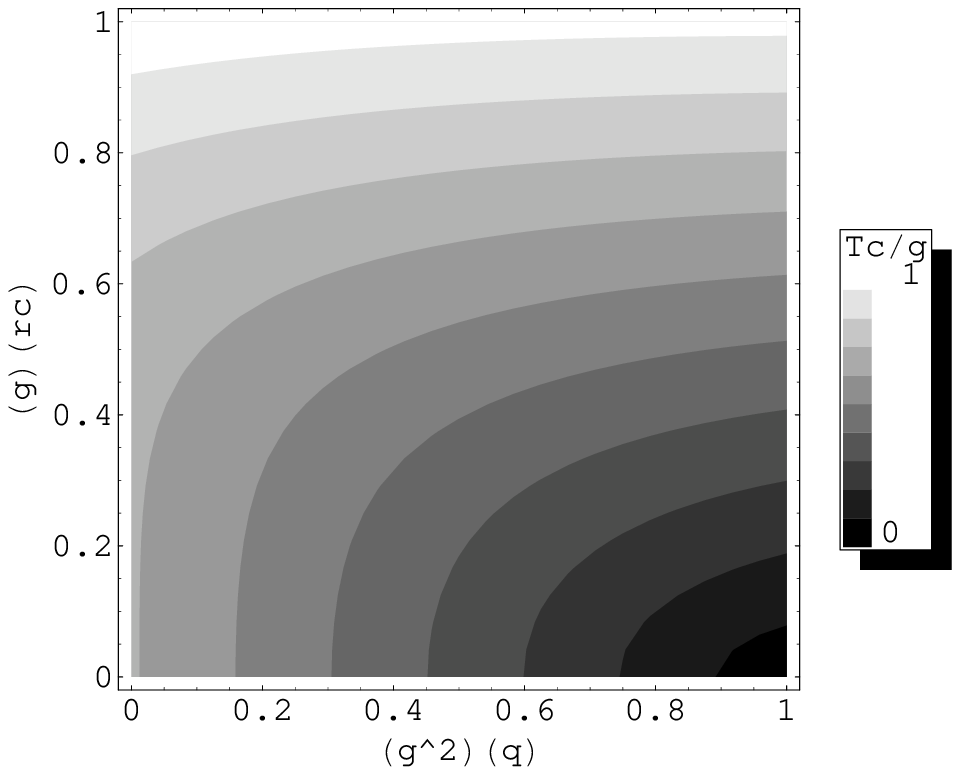}}\\
     \vspace{.3in}
%     \hspace{.1in}
     \subfigure[Hawking-Page analysis of AdS$_7$ R-charged black hole. Only the region where
$gr_c>0.24,g^2q>0$ is shown.  We remark $T_c=\frac{5g}{2\pi}$ for
zero charge density and no IR cut-off.]{
           \label{fig3}
           \includegraphics[width=.45\textwidth]
                {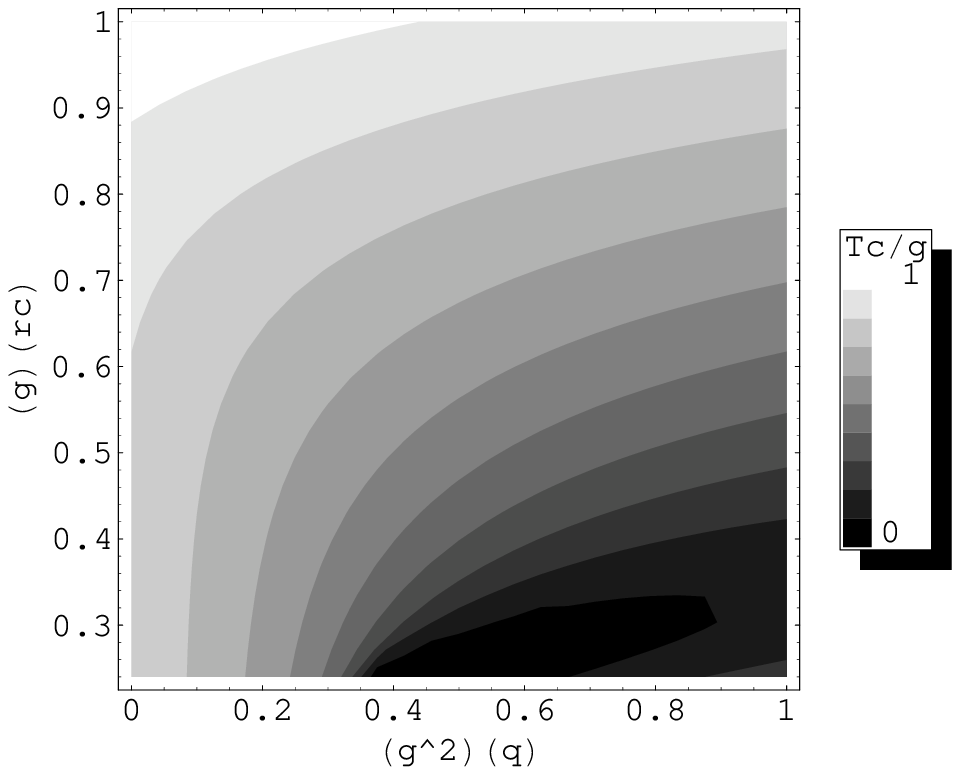}}
\end{flushleft}
     \caption{The contour plot of confinement/deconfinement phase transition temperature $T_c$ as a function of IR
cut-off $r_c$ and charge density $q$.}
\end{figure}

The R-charged black hole solution in four dimensions corresponds
to a particular $U(1)^4$ truncation of maximal gauged supergravity
in $d=4$, which can be traced back to compactification of $d=11$
supergravity. The four charges can be seen as near-extremal black
M$2$ brane spinning along four transverse
directions\cite{Cvetic:1999ne}. The corresponded field theory is
the thermal $d=3,{\cal N}=2$ super Yang-Mills living on the
boundary with topology $R_t\times S^2$. The deconfinement
temperature, as a function of cut-off and charge density, will not
be detailed here for its complicated form, which comes from
solving a third-order polynomial equation of two unknowns.  In the
figure \ref{fig1}, we observe that increasing IR cut-off raises
transition temperature, however, increasing charge density lowers
it.  We find transition temperature $T_c=g/\pi$ for zero charge
density and no IR cut-off.

\subsection{$5d$ black hole/thermal $4d$ super Yang-Mills}
The R-charged black hole solution in five dimensions corresponds
to a particular $U(1)^3$ truncation of maximal gauged supergravity
in $d=5$, which can be traced back to compactification of $d=10$
supergravity on the special geometry, so called STU model.
Therefore the black hole may carry up to three charges, seen as a
near-extremal black D$3$ brane spinning along three transverse
directions\cite{Cvetic:1999ne}. The corresponded field theory is
the thermal $d=4,{\cal N}=2$ super Yang-Mills living on the
boundary with topology $R_t\times S^3$. In the case of single
charge, the transition temperature can be found in a manageable
form,
\begin{eqnarray}
&&g^{-1} T_c =
\frac{3(1+\Delta)-qg^2(4+\Delta)+g^4(q^2+8qr_c^2+8r_c^4)}{\sqrt{2}\pi(1+\Delta-qg^2)\sqrt{1+\Delta+qg^2}},\nonumber\\
&&\Delta\equiv \sqrt{1-2qg^2+g^4(q^2+8qr_c^2+8r_c^4)}.
\end{eqnarray}
It is interesting to compare with the result found in
\cite{Cai:2007zw} at the limit $q\to 0$. In particular, we recover
$T_c=3g/2\pi$ as IR cut-off is also removed.  The regularity of
$\Delta$ and non-negativity of $T_c$ impose constraints on
admissible $q$ and $r_c$.  In the figure \ref{fig2}, we also
observe similar effects of IR cut-off and charge density on the
deconfinement temperature.

\subsection{$6d$ black hole/thermal $5d$ super Yang-Mills}
The R-charged black hole solution in six dimensions corresponds to
a particular $U(1)$ subgroup of $SU(2)\times U(1), {\cal N}=(1,1)$
supergravity in $d=6$\cite{Cvetic:1999xp}.  To be specific, we
have instead two equal charges, i.e.
\begin{equation}
{\cal H}=(1+\frac{q}{r^3})^2.
\end{equation}
According to equation (\ref{V13}),  phase transition is supposed
to happen at
\begin{equation}
\frac{\mu}{2}-r_+^3-g^2r_c^5-2qg^2r_c^2-\frac{q^2g^2}{r_c}=0.
\end{equation}
The last term in the above equation diverges at small $r_c$ for
finite charge density $q$.  Technically, we may set $q=0$ before
sending $r_c\to 0$ to avoid the divergence and render the usual
case of zero charge density and no IR cut-off, where $T_c=2g/\pi$
is expected.  In generic, however, we expect this divergent term
will be cancelled via a counterterm at the IR cut-off as well as
other finite terms might be modified.  Naively dropping this
divergent term, we fail to find another solution beside the
trivial one mentioned above.

\subsection{$7d$ black hole/thermal $6d$ super Yang-Mills}
The R-charged black hole solution in seven dimensions corresponds
to a particular $U(1)^2$ truncation of $SO(5)$ maximal gauged
supergravity, which can be deduced from the $S^4$ reduction of
$d=11$ supergravity.  Therefore the black hole may carry up to two
charges, seen as a near-extremal black M$5$ brane spinning along
two transverse directions\cite{Cvetic:1999ne}.   The corresponded
field theory is the thermal $d=6,{\cal N}=2$ super Yang-Mills
living on the boundary with topology $R_t\times S^5$. The
deconfinement temperature, as a function of cut-off and charge
density, will not be detailed here for its complicated form, which
comes from solving a six-order polynomial equation of two
unknowns. In the figure \ref{fig3}, we find that the regularity of
$\Delta$ and non-negativity of $T_c$ impose more restrict
constraints on admissible $q$ and $r_c$.  Some values of $q$'s are
not allowed for small $r_c$.  However in general, we still observe
similar effects of IR cut-off and charge density on deconfinement
temperature, however, we remark that $T_c$ has minimum around
$g^2q \sim 0.6$ and $gr_c \sim 0.3$.  We find transition
temperature $T_c=5g/2\pi$ for zero charge density and no IR
cut-off.

\section{Conclusion}
In this note we study the confinement/deconfinement phase
transition of gauge theory with fixed chemical potential via the
Hawking-Page analysis of single R-charged black hole in truncated
AdS space of various dimensions.  In the limit of zero charge
density and no IR cut-off, we recover the transition temperature,
\begin{equation}
T_c=(d-2)\frac{g}{\pi},
\end{equation}
for $d$-dimensional AdS black hole or gauge field on $R_t\times
S^{d-2}$, where $4\le d \le 7$.  This is consistent with the known
fact that $T_c\to 0$ as $g\to 0$.  For nonzero charge density $q$
and IR cut-off $r_c$, we obtain $T_c$ as a function of $q$ and
$r_c$ for each case except $d=6$, where the difference of Gibbs
free energy diverges at $r_c\to 0$.  We expect that a counterterm
constructed at this IR boundary in the way similar to those in
\cite{Batrachenko:2004fd} will render it finite, and then the
improved function $T_c$ will make better sense.  We leave a
comprehensive treatment for future projects.  In the dimension
$d=4,5,7$, $T_c$ in general decreases with increasing $q$ but
decreasing $r_c$. Similar relation between $T_c$ and $q$ was also
observed in \cite{Kim:2007em} though their matter field came from
a different source.  This relation might be understood as follows:
At first, high charge density makes {\it quarks} (charged under
global $U(1)$) expel from each other, weakening the confining
force and thus lowering the confinement/deconfinment transition
temperature. Secondly, Hawking-Page transition happens whenever
the black hole horizon $r_+$ is around IR cut-off $r_c$. From
equation (\ref{temp}) we learn that temperature $T_c$ grows with
$r_+$. Therefore we can deduce that $T_c$ decreases with
decreasing $r_c$.  However, we notice that in the case of $d=7$,
$T_c$ bounces back for high charge density.  Though this seems a
contradiction to our previous statement, a possible resolution can
be given as follows:  Once $T_c$ has reached its minimum, charge
density also reaches its critical value of saturation.  If one
tries to put more {\it quarks} in the system, any pair of them is
forced to recombine by expelling from the surrounded others. Each
pair takes extra energy to be deconfined and therefore $T_c$
raises up.  As the last remark, the author in \cite{Herzog:2006ra}
showed that a more realistic meson spectrum can be produced via
the soft wall model rather than the hard one, it might be
interesting to repeat this Hawking-Page analysis in the soft wall
model of the same R-charged black hole background.

\begin{acknowledgments}
The author would like to thank Furuuchi Kazuyuki for useful
discussion.  This work is supported in part by the Taiwan's
National Science Council under Grant No. NSC96-2811-M-002-018.
\end{acknowledgments}

\end{document}